\documentclass[confere-eps-converted-to.pdfnce, compsoc, onecolumn, 11pt, letter]{IEEEtran}

\usepackage{amsfonts}
\usepackage{cite}
\usepackage{graphicx}
\usepackage{epsfig}
\usepackage[cmex10]{amsmath}

\newcommand{\squishlist}{
 \begin{list}{$\bullet$}
  { \setlength{\itemsep}{0pt}
     \setlength{\parsep}{3pt}
     \setlength{\topsep}{3pt}
     \setlength{\partopsep}{0pt}
     \setlength{\leftmargin}{1.5em}
     \setlength{\labelwidth}{1em}
     \setlength{\labelsep}{0.5em} }}

\newcommand{\squishlisttwo}{
 \begin{list}{$\bullet$}
  { \setlength{\itemsep}{0pt}
     \setlength{\parsep}{0pt}
    \setlength{\topsep}{0pt}
    \setlength{\partopsep}{0pt}
    \setlength{\leftmargin}{2em}
    \setlength{\labelwidth}{1.5em}
    \setlength{\labelsep}{0.5em}}}  

\newcommand{\squishend}{
  \end{list}  }

\IEEEoverridecommandlockouts

\begin{document}
\title{Tiny Descriptors for Image Retrieval with Unsupervised Triplet Hashing}

\author{
\IEEEauthorblockN{Jie Lin$^{1}$$^{\ast}$, Olivier Mor\`ere$^{1,2}$$^{\ast}$, Julie Petta$^{3}$$^{\ast}$, Vijay Chandrasekhar$^{1}$, Antoine Veillard$^{2}$}
\IEEEauthorblockA{
	$^1$Institute for Infocomm Research, Singapore\\
	$^2$Universit\'e Pierre et Marie Curie, Paris, France\\
	$^3$CentraleSup\'elec, Paris, France\\
}
\thanks{$^{\ast}$ Equal contributions from Jie Lin, Olivier Mor\`ere and Julie Petta.}
}

\maketitle

\begin{abstract}

A typical image retrieval pipeline starts with the comparison of global descriptors from a large database to find a short list of candidate matches.
A good image descriptor is key to the retrieval pipeline and should reconcile two contradictory requirements: 
providing recall rates as high as possible and being as compact as possible for fast matching.
Following the recent successes of Deep Convolutional Neural Networks (DCNN) for large scale image classification, 
descriptors extracted from DCNNs are increasingly used in place of the traditional hand crafted descriptors such as Fisher Vectors (FV) with better retrieval performances.
Nevertheless, the dimensionality of a typical DCNN descriptor --extracted either from the visual feature pyramid or the fully-connected layers-- 
remains quite high at several thousands of scalar values.

In this paper, we propose Unsupervised Triplet Hashing (UTH), a \emph{fully unsupervised} method to compute extremely compact binary hashes 
--in the 32-256 bits range-- from high-dimensional global descriptors.
UTH consists of two successive deep learning steps.
First, Stacked Restricted Boltzmann Machines (SRBM), a type of unsupervised deep neural nets, 
are used to learn binary embedding functions able to bring the descriptor size down to the desired bitrate.
SRBMs are typically able to ensure a very high compression rate at the expense of loosing some desirable metric properties of the original DCNN descriptor space.
Then, triplet networks, a rank learning scheme based on weight sharing nets is used to fine-tune the binary embedding functions to retain as much as possible of the useful metric properties of the original space.
A thorough empirical evaluation conducted on multiple publicly available dataset using DCNN descriptors shows that our method is able to significantly outperform state-of-the-art unsupervised schemes in the target bit range.

\end{abstract}

\section{Introduction}

For mobile visual search and augmented reality applications, the size of data sent over the wireless network needs to be as
small as possible to reduce latency and improve user experience. 
One approach to the problem is to transmit JPEG compressed images or MPEG compressed videos over the wireless network, 
but this might be prohibitively expensive at low uplink speeds. 
An alternate approach to sending JPEG images or MPEG videos is to extract feature descriptors on the mobile device, 
compress the descriptors and transmit them over the wireless network. 
Such an approach has been demonstrated to reduce the amount of transmission data by orders of magnitude 
for both visual search and augmented reality applications~\cite{CHoGJournal}~\cite{CHoGVideo}. 
To this end, MPEG has issued an international standard titled Compact Descriptors for Visual Search (CDVS)~\cite{CDVS1} for descriptor extraction and compression very recently, 
and is embarking on extending the CDVS standard to video sources, titled Compact Descriptors for Video Analysis (CDVA).

State-of-the-art content-based image retrieval pipelines consist of two blocks: (1) retrieving a subset of images from the database that are similar, 
and (2) using Geometric Consistency Checks (GCC) (e.g., RANSAC) for finding relevant database images with high precision. 
The GCC step is computationally complex and can only be performed on a small number of images (hundreds). 
As a result, the first step of the pipeline is critical to achieving high recall.
For the first step, state-of-the-art schemes are based on comparing global representations of images.  
The {\it global descriptor} of an image is represented by a single high dimensional vector with tens of thousands of dimensions.
Examples include Fisher Vectors~\cite{Perronnin_CVPR_10}, Residual Enhanced Visual Vector(REVV)~\cite{REVV1}, 
Scalable Compressed Fisher Vector (SCFV)~\cite{RCFC}, and the recently proposed descriptor based on Deep Convolutional Neural Networks (DCNN)~\cite{Yandex}, 
such as AlexNet~\cite{AlexNet} and OxfordNet~\cite{VeryDeepNeuralNets}.
Subsequently, {\it local descriptors} like SIFT~\cite{Lowe99} and CHoG~\cite{CHoGJournal} are used in the GCC step to check 
if a valid geometric transform exists between database and query images.

The problem of {\it global} descriptor compression is an important one.
The more compact the {\it global} descriptor, the faster the descriptors can be compared for matching.
Further, it is highly desirable that the global descriptors be binary to enable fast matching through Hamming distances.
With internet-scale image databases, like the recently released Yahoo 100M image database~\cite{Yahoo100MDataset}, 
compact global descriptors will be key to fast web-scale image-retrieval.
Ideally, a 32-bit or 64-bit binary global descriptor is highly desirable, as it can be directly addressed in RAM.
However, finding such a representation is extremely challenging.

\section{Related Work and Contributions}

Motivated by the remarkable success of deep learning for large-scale image classification~\cite{AlexNet}~\cite{VeryDeepNeuralNets}, 
recent work show that the use of representations extracted from DCNN is quickly gaining ground over hand-crafted descriptors 
such as Fisher Vectors as favoured state-of-the-art global image descriptors for image instance retrieval~\cite{Yandex}~\cite{EvaluationCNN}.
While neural networks have been around for several decades, their resurgence can be attributed to two key factors: 
availability of large training data, and large amounts of computing power, which makes training large and deep networks possible.
E.g., the neural networks in~\cite{AlexNet}~\cite{VeryDeepNeuralNets} have 7 and 19 layers respectively, 
and take weeks to train with millions of images on GPU.

While there is plenty of work on learning binary codes~\cite{KristenHashingSurvey} for compressing small descriptors like SIFT,
there is relatively little work on compression of high-dimensional global descriptors.
One promising approach is unsupervised hashing.
The goal of unsupervised hashing is to compress original descriptors into small binary codes with binary embedding functions.
These binary embedding functions are usually learned from data without any side information.
For example,
Gong et al. proposed the popular Iterative Quantization (ITQ)~\cite{ITQ}.
ITQ first performs PCA and then learn a rotation to minimize the quantization error of mapping the transformed data to the vertices of a zero-centered binary hypercube. 
Our previous work~\cite{CompactGlobal} employed deeply stacked Restriction Boltzmann Machines (SRBM) to learn low dimensional non-linear subspaces of uncompressed descriptors.
Other examples include Locality Sensitive Hashing (LSH)\cite{LSH}, Spectral Hashing (SH)~\cite{SpectralHashing} and Bilinear Projection Binary Codes (BPBC)~\cite{BPBC}, etc.
We note that most of the unsupervised hashing schemes formulate image retrieval as an optimization problem,
following either the principle of Mean Squared Error (MSE) like ITQ or Maximum-Likelihood Estimation (MLE) like SRBM.
These hashing schemes are sub-optimal solutions because they ignore ranking order information in image retrieval,
i.e., relevant images should be ranked ahead of irrelevant ones for queries.

Besides hashing schemes, quantization based method such as Product Quantization (PQ)~\cite{PQFisher} are an alternative way to compress descriptors.
Jegou et al. proposed PCA followed by random rotations and PQ for obtaining compact representations~\cite{PQFisher}. 
While this results in highly compact descriptors, the resulting representation is not binary and cannot be compared with Hamming distances.
The MPEG CDVS standard adopted the Scalable Compressed Fisher Vector~\cite{RCFC}, which was based on direct scalar quantization of high-dimensional Fisher Vectors.  
The size of the compressed descriptor in the MPEG-CDVS standard ranges from 256 bytes to several thousand bytes per image, based on the operating points.
In this paper, we focus on extremely low bitrate binary coding (32 to 256 bits) of high-dimensional global descriptors, 
thus, the quantization approaches are outside the scope of this work.

We propose Unsupervised Triplet Hashing (UTH), a scheme for learning binary embedding functions of high-dimensional representations.
UTH consists of a two-stage deep learning pipeline.
First, we use SRBMs to learn a first version of the binary embedding functions in a fashion similar to our previous work in~\cite{CompactGlobal} where we showed that the method is able to produce very compact hashes with good retrieval performances.
However, SRBMs is purely based on data reconstruction and do not purposefully try to preserve the good metric properties of the original high-dimensional space.
Therefore, we add a second step to fine-tune the network weights (embedding parameters).
This second step uses triplets of weight-sharing networks and learns to preserve the ranking order of triplets of images.
Unlike other approaches using triplet learning networks~\cite{triplet}~\cite{tripletrank}~\cite{triplet2}, our approach is \emph{fully-unsupervised} and does not require additional label data for the triplets.
For our experiments, we use for our starting representation an intermediate layer from the 19-layer OxfordNet which is state-of-the-art for image classification.

Our contributions are three-fold:
\squishlist
\item We propose a method for learning binary embedding functions able to produce very compact hashes (down to 32 bits) for image instance retrieval.
UTH expands our previous work on hash compression~\cite{CompactGlobal} with a fine-tuning step based on triplet networks to preserve the ranking information 
from the high-dimensional global descriptors.
\item Unlike with other approaches, the learning-to-rank pipeline is fully unsupervised and does not require any additional label data.
To our knowledge, this is the first work on learning to rank with unsupervised deep network applied to image instance retrieval.
\item Through a thorough empirical evaluation on small and large datasets, we show that UTH can reduce the data size of uncompressed descriptors by 512$\times$ (256 bits hashes) without considerable retrieval performance loss.
We also show that UTH is able to outperform other unsupervised schemes in the 32-256 bits range.
\squishend

\section{Evaluation Framework}
\label{sec:eval}

We use 3 popular data sets for small scale experiments: 
INRIA Holidays (500 queries, 991 database images)~\cite{Jegou08}, 
University of Kentucky Benchmark (UKbench) (10200 queries, 10200 database images)~\cite{Nister06}, 
and Oxford5k Buildings (55 queries, 5063 database images)~\cite{Philbin07}. 
For large-scale retrieval experiments, 
we present results on Holidays and UKbench data sets, combined with the 1 million MIR-FLICKR distractor data set~\cite{mirflickr}.
Note that the UKbench dataset is used in the CDVS evaluation framework~\cite{CDVS1}.

Most schemes, including our proposed scheme, require a training step. 
We use the ImageNet data set for training, which consists of 1 million images from 1000
different classes~\cite{DengImagenet} (class labels are never used in the scope of this work).
We randomly sample a subset of 150K images from ImageNet to learn binary embedding functions.
This training set is independent of the query and database data described above ensuring there is no over-fitting while testing.

The starting global image descriptor is extracted from the 19-layer OxfordNet proposed for ImageNet classification 
in the seminal contribution by the VGG group~\cite{VeryDeepNeuralNets}.
The features are extracted using the open-source software Caffe~\cite{Caffe}. 
We find that layer fc6 (the first fully connected layer before softmax) performs the best for image retrieval,
similar to the recently reported results in~\cite{Yandex}. 
We refer to this 4096-dimensional fc6 features as the DCNN features from here-on.

\begin{figure*}
	\centering{
		\begin{tabular}{@{}c@{} @{}c@{} @{}c@{} }
			\includegraphics[width=2.3in,height=1.65in]{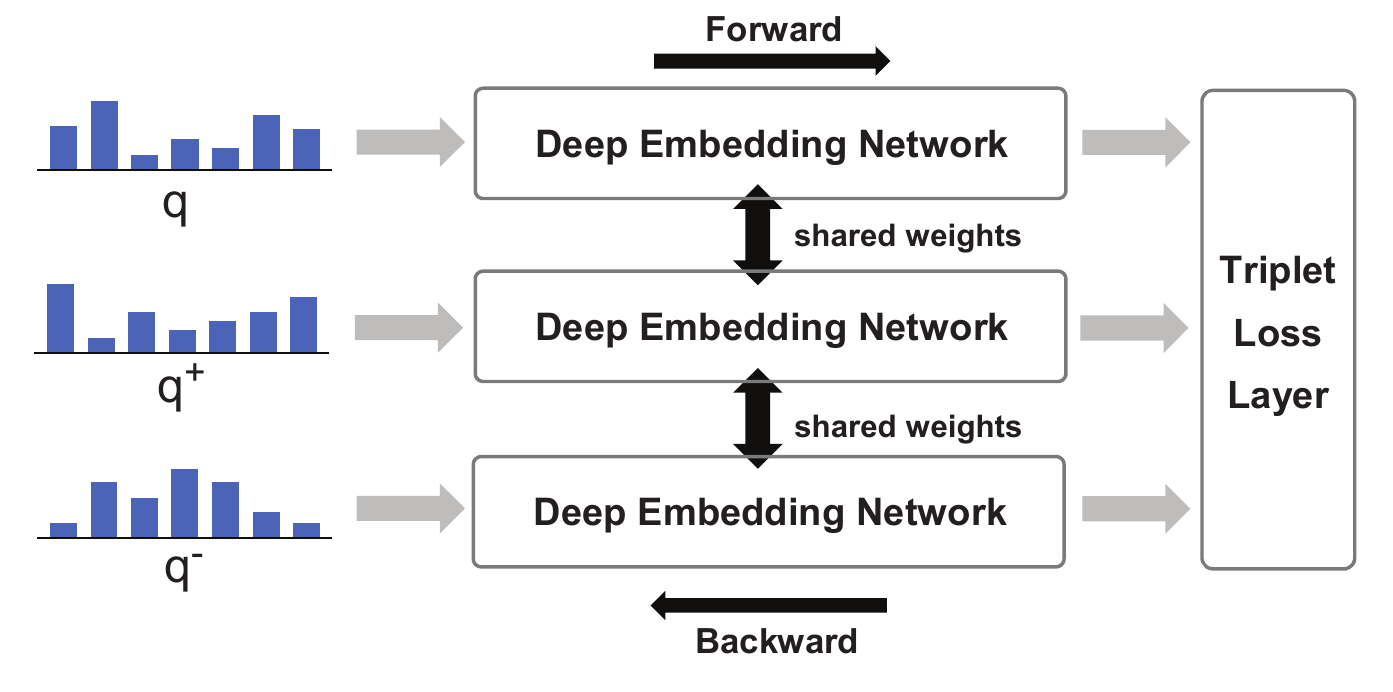} &
			\includegraphics[width=1.65in,,height=1.65in]{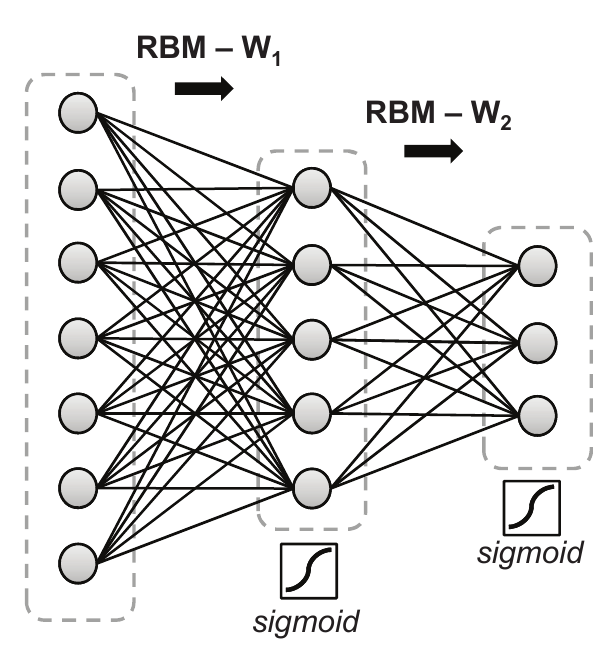} &
			\includegraphics[width=1.8in,,height=1.8in]{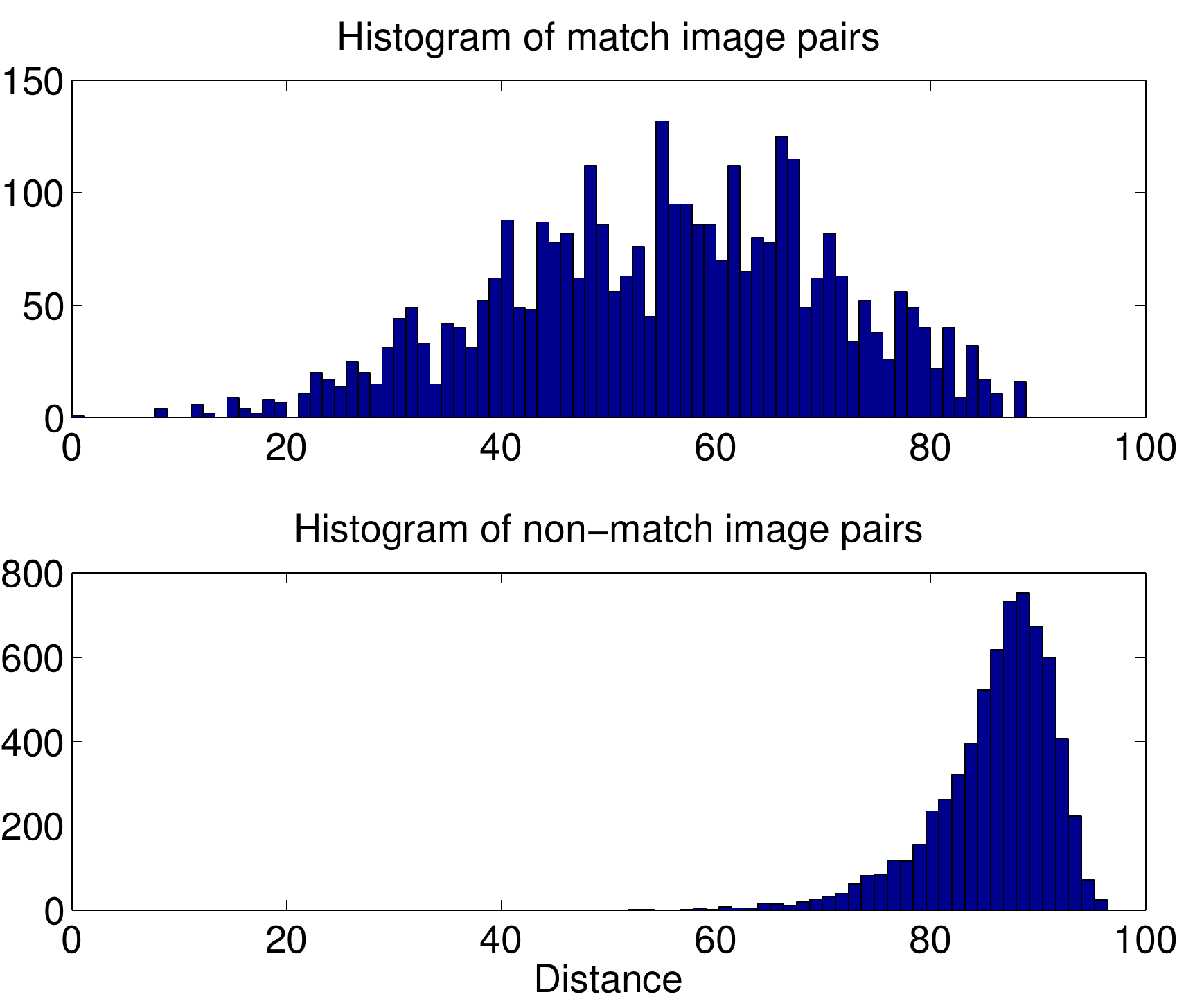} \\
			{\it (a)} & {\it (b)} & {\it (c)} \\
		\end{tabular}
		\caption{\footnotesize (a) The proposed unsupervised triplet hashing scheme; 
	        (b) The Deep Embedding Network are pre-trained using Stacked RBMs (SRBM); 
			(c) Distribution of squared Euclidean distances for DCNN descriptors from $3,300$ match image pairs and $6,600$ non-match image pairs. 
			It shows that the original uncompressed descriptors already contain good ranking information.}
		\label{fig:structure}
		}	
\end{figure*}

\section{Unsupervised Triplet Hashing}
\label{sec:rbm}

This work focuses on compressing high-dimensional descriptors into low bitrate binary codes for image instance retrieval.
UTH is a two-step procedure.
First, we use SRBMs to pre-train an embedding model able to achieve the desired bitrate.
Next, we fine-tune the model in order to preserve the ranking information from the original uncompressed descriptors using triplets of images.
The entire process is unsupervised and does not require any labeled data and is illustrated in Figure~\ref{fig:structure}.

{\bf Deep Embedding Network.}
Our previous work showed that high dimensional vectors can be converted to low-dimensional codes by 
training multi-layer neural networks based on stacked Restricted Boltzmann Machines (SRBM), 
which can perform significantly better than most unsupervised hashing approaches for image instance retrieval~\cite{CompactGlobal}.

An RBM is an undirected bipartite graphical model consisting of a layer of visible (input) units $v$ and a layer of hidden (output) units $h$.
A set of symmetric weights $W$ connects $v$ and $h$. 
For an RBM with visible and hidden binary units,
the activation probabilities of units in one layer can be sampled by fixing the states of the other layer as follows:
\begin{equation}
\mathbb{P}(h_j = 1 | v) = \sigma (b_j + \sum_i w_{ij} v_i )
\label{eq:alt_1}
\end{equation}
\begin{equation}
\mathbb{P}(v_i = 1 | h) = \sigma (b_i + \sum_j w_{ij} h_j )
\label{eq:alt_2}
\end{equation}
where $v_i$ and $h_j$ are the binary states of visible and hidden units $i$ and $j$ respectively, 
$w_{ij}$ are the weights connecting the units, $b_i$ and $b_j$ are their respective bias terms,
and $\sigma(\cdot)$ is the sigmoid function.
RBM can be trained by minimizing the contrastive divergence objective~\cite{RBMPracticalGuide}, which approximates the maximum likelihood of the input distribution.
Alternating Gibbs sampling based on Equations~(\ref{eq:alt_1}) and (\ref{eq:alt_2}) is used to obtain the network states 
to update the parameters $w_{ij}, b_i, b_j$ through gradient descent.

We propose stacking multiple RBMs (SRBM) to create a deep embedding network with several layers. 
Each layer captures higher order correlations between the units of the previous layer in the network.
An SRBM model with two stacked RBMs is illustrated in Figure~\ref{fig:structure} (b).
The first layer of the deep embedding network takes a high-dimensional image descriptor $q$ as visible input.
We use binary latent units with sigmoid activation function, because binary output bits are desired for our hash.
Besides, binary RBMs are a lot faster and easier to train than continuous RBMs~\cite{RBMPracticalGuide}.
We perform greedy layer-by-layer training by fully training one RBM at a time using contrastive divergence. 
Each new RBM layer models the output layer of the previous layer.
The number of layers and the number of hidden units in each layer are parameters that are typically chosen experimentally~\cite{HintonScience}~\cite{AlexNet}~\cite{VeryDeepNeuralNets}. 
Here, we progressively decrease the dimensionality of hidden layers by a factor of 2, 
and train several RBMs with varying number of hidden layers and output units to optimize parameters.
More details are available in our previous work~\cite{CompactGlobal}.

While great at achieving high compression rates, SRBMs do not take into account the metric properties of the high-dimensional input space which is not ideal for image instance retrieval.
Accordingly, we perform model fine-tuning in the next step aimed at preserving the good retrieval properties of the original uncompressed descriptors throughout the dimensionality reduction pipeline.

{\bf Triplet Networks.}
The models obtained from the previous step are fine-tuned using the ranking information provided through triplets.
A triplet $(q,q^{+},q^{-})$ contains a query image descriptor $q$, a positive image descriptor $q^{+}$ and a negative image descriptor $q^{-}$,
query $q$ is more similar (closer) to positive image $q^{+}$ than to negative image $q^{-}$.
We learn a binary embedding function $p : \mathbb{R}^n \mapsto 2^{m} $, typically $n >> m$,
such that $d(p(q),p(q^{+})) < d(p(q),p(q^{-}))$.
Accordingly, we define a triplet ranking loss, $l(q,q^{+},q^{-})= max\{0, g+d(p(q),p(q^{+}))-d(p(q),p(q^{-}))\}$,
where $g$ is a positive margin parameter.
By normalising the two distances with softmax (denoted $d'$) and taking $g=1$, we can rewrite the triplet loss function as,
\begin{equation}
l(q,q^{+},q^{-})= max\{0, 1+d'(p(q),p(q^{+}))-d'(p(q),p(q^{-}))\} 
\label{eq:triplet_loss_e}
\end{equation}

The idea of using weight sharing network for model fine-tuning is not new and previous work can be found for image classification and semantic retrieval~\cite{triplet}~\cite{tripletrank}~\cite{triplet2}.
Unlike with previously proposed approaches, triplet learning in UTH in fully unsupervised and does not require dedicated labeled data.
Based on the observation that the original uncompressed descriptors already provide good retrieval performance, we simply construct triplets according to the Euclidean distance in the original space, i.e. such that $\lVert q-q^{+} \rVert_2 < \lVert q-q^{-} \rVert_2$.
Accordingly, our global objective function is,
\begin{equation}
min \quad l(q,q^{+},q^{-}) \quad
s.t.\ \lVert q-q^{+} \rVert_2 < \lVert q-q^{-} \rVert_2
\label{eq:objective_fun}
\end{equation}
The objective function is solved by Stochastic Gradient Descent (SGD)~\cite{triplet}.

{\bf Triplet Sampling.}
A simple sampling strategy is to take three random images, arbitrarily chose $q$ and label two of them as $q^{+}$ and $q^{-}$ according to their relative distance to $q$.
However, we should note that in the context of image retrieval, we are only interested in correctly discriminating ranks for the high end of the retrieval list.
The space of all possible random triplets is also very vast.
Accordingly, we need a more targeted sampling strategy.

We propose a threshold sampling method to generate informative triplets.
Given a training set containing $M$ images, we build a loop-up table with $M$ buckets offline.
For the $m$-th bucket, we compute distances between the $m$-th image descriptor and the rest,
rank the distances in descending order and store the distances and their associated image IDs in that bucket.
To sample a triplet, we first randomly draw an image $q$ from the training set, 
then we sample a positive image $q^{+}$ from the bucket associated with $q$, 
with the constraint that $\lVert q-q^{+} \rVert_2$ should approach to a pre-defined threshold $T_p$.
Similarly, we sample a negative image $q^{-}$ from the same bucket such that $\lVert q-q^{-} \rVert_2 \to T_n$ .
We note that $T_p<T_n$.
To avoid over-fitting, a large number of informative triplets would be required in the fine-tuning step.
In our case, we found the optimization converges to a sweet point with a number of 128K triplets for each epoch.

{\bf Generate binary codes for new image.}
We use the fine-tuned deep embedding network to compress the high-dimensional image descriptors into binary codes.
As shown in Figure~\ref{fig:structure} (b), 
each output component in the final layer is obtained by a composite of several non-linear functions (feedforward projection),
followed by component-wise binarization at 0.5 to produce the binary codes,
\begin{equation}
b=
\begin{cases}
    1, & \text{if } g_i>0.5 \\
    0, & \text{otherwise}
\end{cases}
\end{equation}
where $g_i$ denotes the $i$-th component of the deep network output $\sigma ( b^2 + \sum w^2 \sigma (b^1 + w^1 q) )$ in Figure~\ref{fig:structure} (b),
$w^l$ and $b^l$ are fine-tuned weights and bias terms for $l$-th layer, respectively.

\begin{figure*}
	\centering{
		\begin{tabular}{@{}c@{} @{}c@{} @{}c@{} }
		    \includegraphics[width=1.88in]{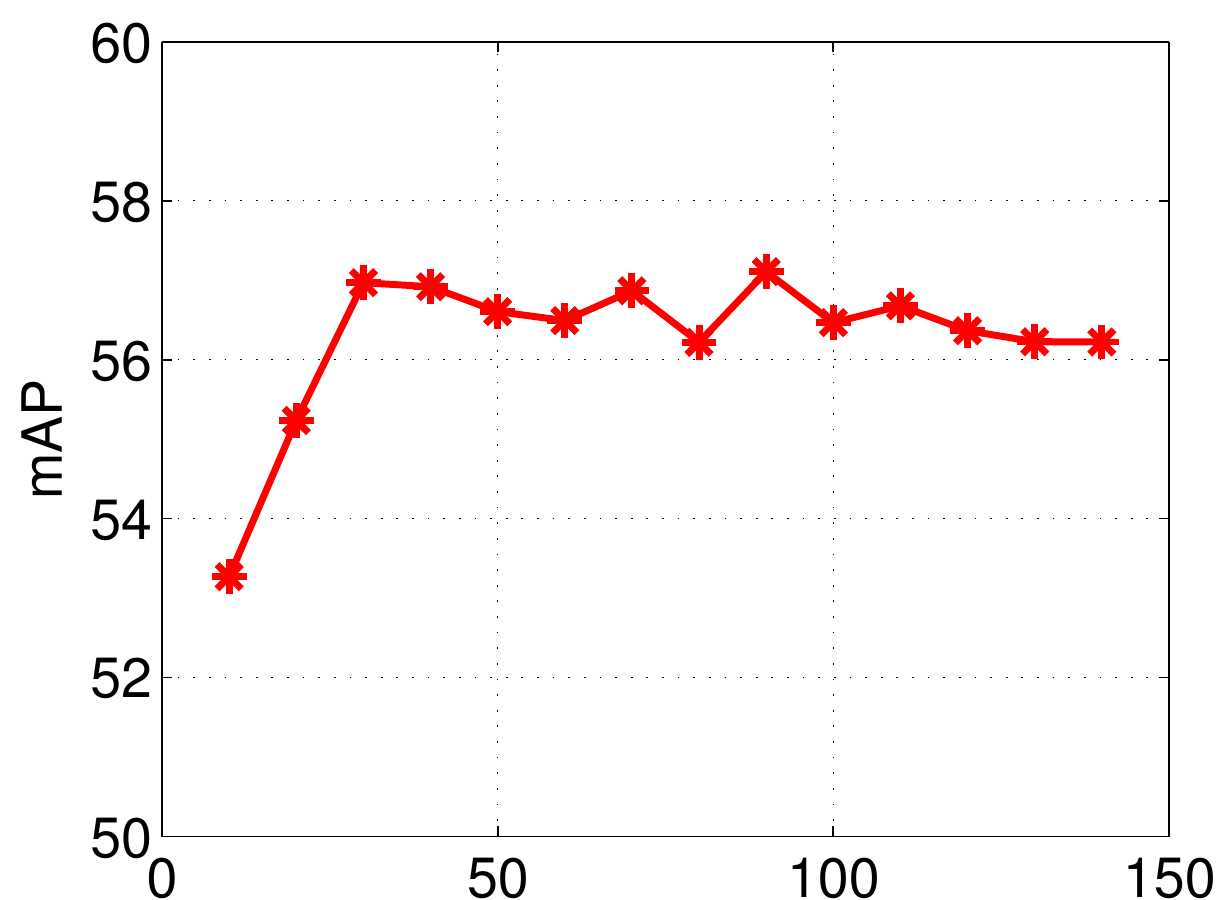} &
		    \includegraphics[width=2in]{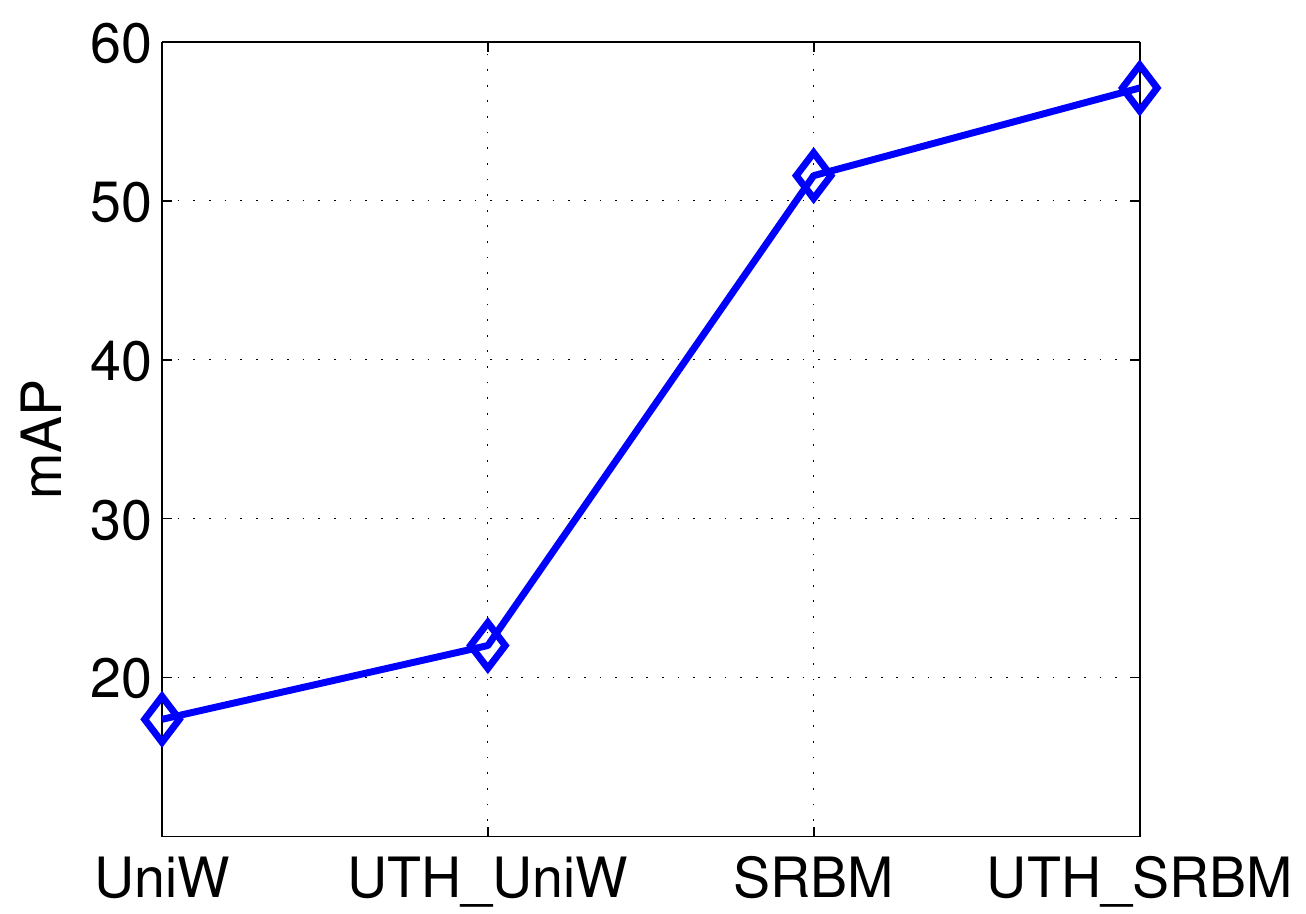} &
		    \includegraphics[width=1.84in]{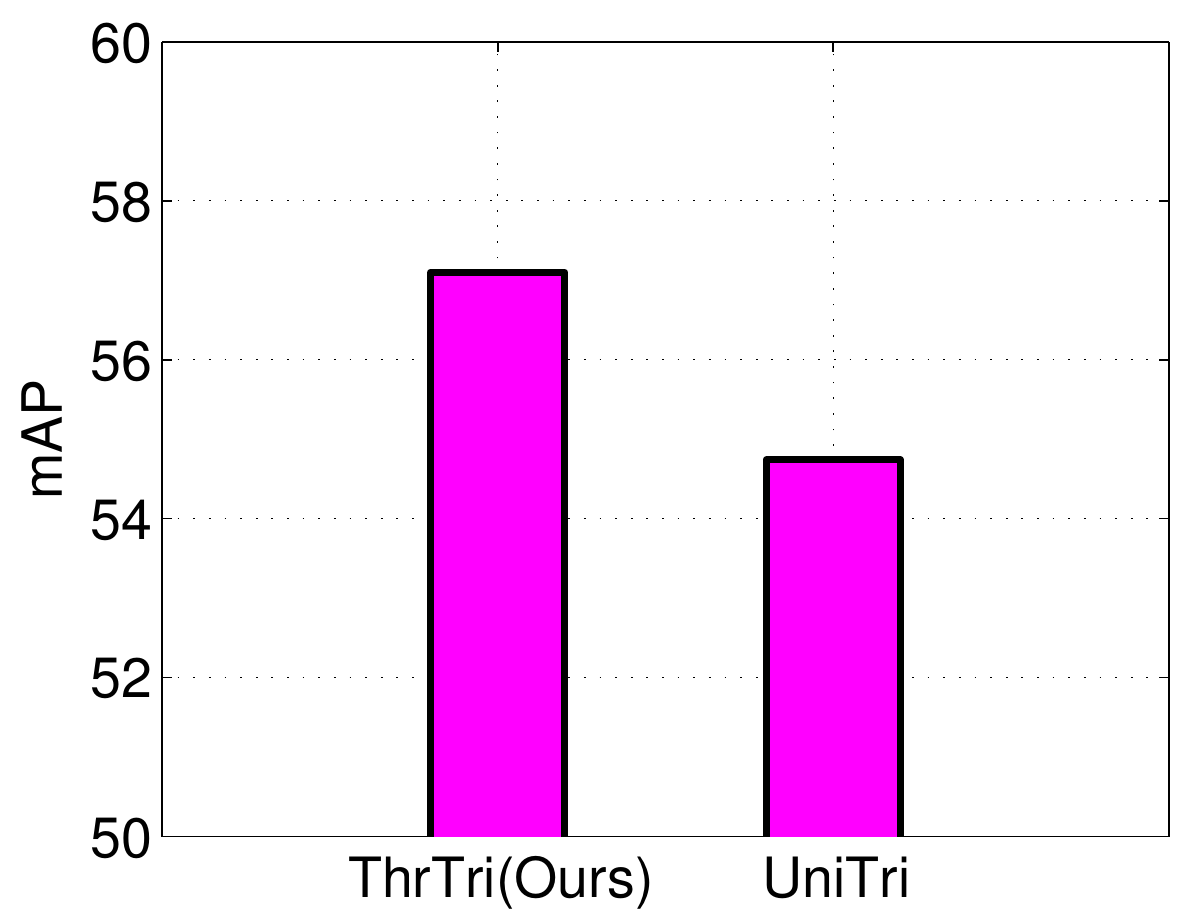} \\
			{\it (a) training epochs} & {\it (b) network weights initilization} & {\it (c) triplet sampling} \\
		\end{tabular}
		\caption{\footnotesize (a) The effect of training epochs in the fine-tuning stage based on unsupervised triplet hashing; 
			(b) Performance comparisons of pre-trained network weights with SRBM vs. random generated unit-norm network weights, with or without fine-tuning (UTH);
			(c) Performance comparisons of the proposed threshold triplet sampling (ThiTri) vs. uniformly sampled triplets (UniTri).
			All results are reported in terms of mAP on Holidays dataset at $64$ bits.}
		\label{fig:performance_effect}
		}	
\end{figure*}

\section{Experimental Results}
\label{sec:exp}

We conduct experiments for the following extremely low bit rates: $32$, $64$, $128$ and $256$ bits.
For the proposed UTH scheme, 
we use the following SRBM layer sizes: $4K-2K-256$, $4K-2K-128$, $4K-1K-64$ and $4K-2K-32$ for $\{256, 128, 64, 32\}$ bits respectively. 
These parameters are chosen from the greedy optimization discussed in our previous work~\cite{CompactGlobal}.
During training, we set the learning rate in the range $[0.0005, 0.01]$ for the weight and bias parameters, momentum to 0.9, 
and ran the training for a maximum 150 epochs.
We illustrate the impact of fine-tuning epochs in terms of mAP on Holidays in Figure~\ref{fig:performance_effect}(a),
the results show that the triplet loss optimization in the fine-tuning stage converges quickly after a few epochs.
Next, we evaluate the key components of the proposed scheme 1) SRBM based network weights initialization in the pre-training stage and 
2) triplet sampling in the fine-tuning stage.

To evaluate the effect of network weights initialization in the pre-training stage,
we present mAP results on {\it Holidays} dataset at output size $64$ bits (see Figure~\ref{fig:performance_effect}(b)),
for comparison (1) SRBM based network weights proposed in our previous work~\cite{CompactGlobal} with (2) random unit-norm network weights (denoted as UniW).
In addition, we report the results combining the proposed UTH scheme with either SRBM (denoted as UTH\_SRBM) or UniW (denoted as UTH\_UniW).
Firstly, SRBM largely improves mAP from 17.4$\%$ to 51.6$\%$, compared to UniW. 
It shows that pre-training the deep network properly is important.
Secondly, our UTH can further boost the performance when combined with SRBM (+5.5$\%$ for UTH\_SRBM) or UniW (+4.6$\%$ for UTH\_UniW).
This demonstrates the effectiveness of UTH in the fine-tuning stage.

To evaluate the impact of triplet sampling strategies in the fine-tuning stage,
we report the results in terms of mAP on Holidays at $64$ bits, 
for comparison (1) the proposed threshold triplet sampling (denoted as ThrTri) with (2) uniformly triplet sampling (denoted as UniTri).
Note that the number of triplet samples used for fine-tuning is the same for both sampling methods.
As shown in Figure~\ref{fig:performance_effect}(c), 
we observe that our ThrTri performs significantly better than UniTri (54.7$\%$ vs. 57.1$\%$).
This result shows the proposed triplet sampling method can choose more informative triplets for fine-tuning.

Next, we compare retrieval experiments for several state-of-the-art hashing and compression schemes. 
Some of these schemes have been proposed for lower dimensional vectors like SIFT, 
but we evaluate their performance on high-dimensional DCNN features.
(1) {\it LSH}~\cite{LSH}.
LSH is based on random unit-norm projections of the DCNN, followed by signed binarization.
(2) {\it SKLSH}~\cite{SKLSH}.
Shift-invariant Kernel LSH (SKLSH) is a distribution-free encoding scheme based on random projections,
such that the expected Hamming distance between the binary codes of two vectors is related to the value of a shift-invariant kernel between the vectors.
(3) {\it SH}~\cite{SpectralHashing}.
Spectral Hashing (SH) is to minimize the sum of the Hamming distances between pairs of binary codes weighted by the Gaussian kernel between the corresponding vectors.
(4) {\it PCAHash}~\cite{ITQ}. 
PCA is applied to the data, and the top ranked dimensions are retained.
A random rotation matrix is then applied to balance the variance across projected dimensions, followed by signed binarization.
(5) {\it ITQ}~\cite{ITQ}. 
For the Iterative Quantization (ITQ) scheme, the authors proposed signed binarization after applying two transforms: 
first the PCA matrix, followed by a rotation matrix, 
which minimizes the quantization error of mapping the PCA-transformed data to the vertices of a zero-centered binary hypercube.
(6) {\it BPBC}~\cite{BPBC}.
For high dimensional data, the PCA projection matrix might require hundreds of megabytes stored in memory. 
Instead of the large projection matrices used in~\cite{ITQ}, the authors apply bilinear random projections, 
which require far less memory, to transform the data. 
This is followed by signed binarization to generate binary hashes.
(7) {\it SRBM}~\cite{CompactGlobal}.
SRBM is the unsupervised deep hashing scheme based on stacked RBMs, as proposed in our previous work.
(8) {\it UTH}.
The proposed unsupervised triplet hashing (UTH) scheme, fine-tune the deep network initialized by SRBM.

\begin{figure*}
    \includegraphics[width=5.7in]{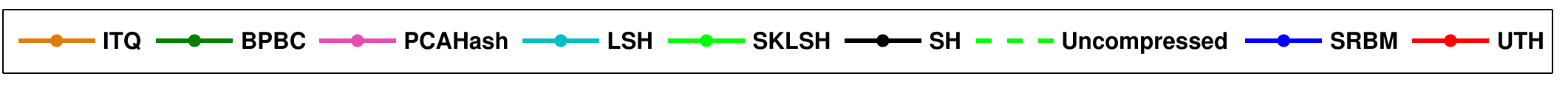}
	\centering{
		\begin{tabular}{@{}c@{} @{}c@{} @{}c@{}}
			\includegraphics[width=2in]{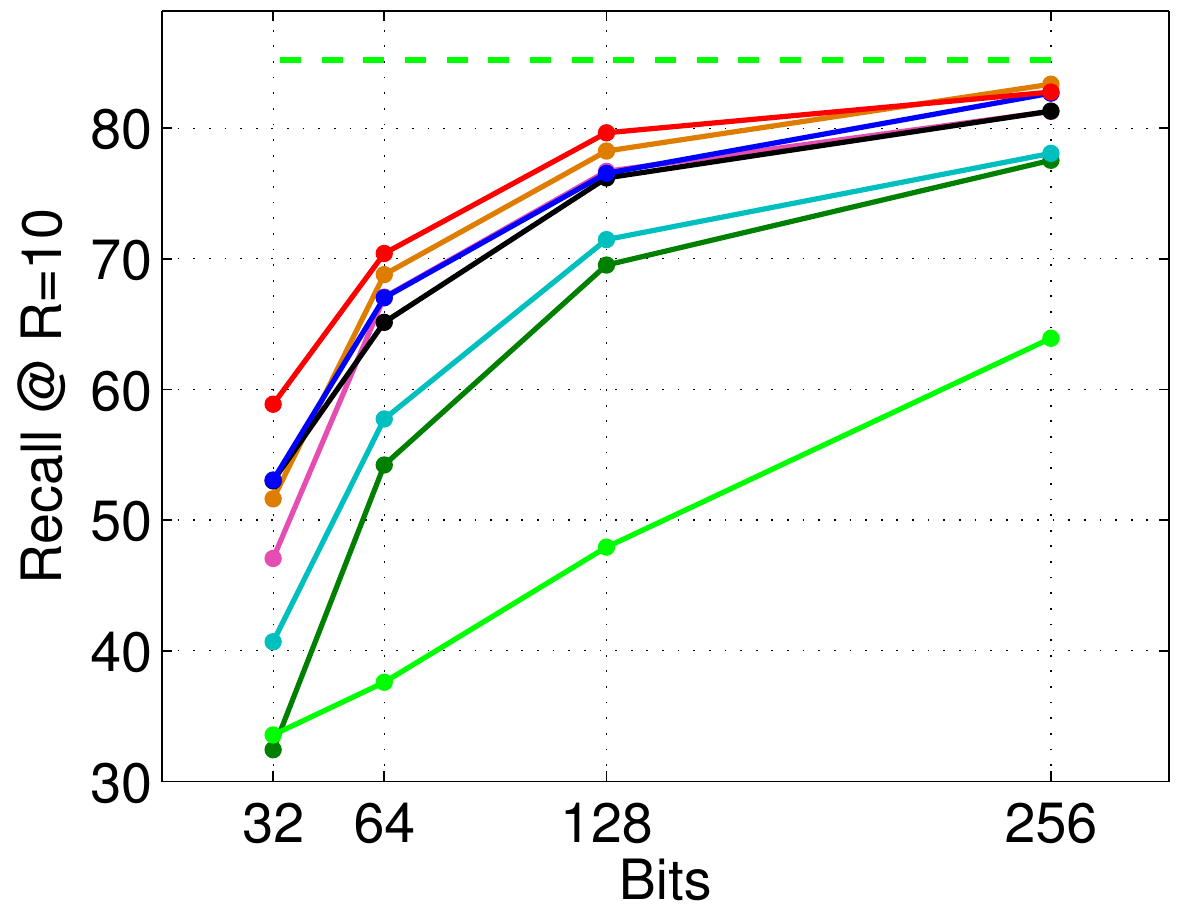} &
			\includegraphics[width=2in]{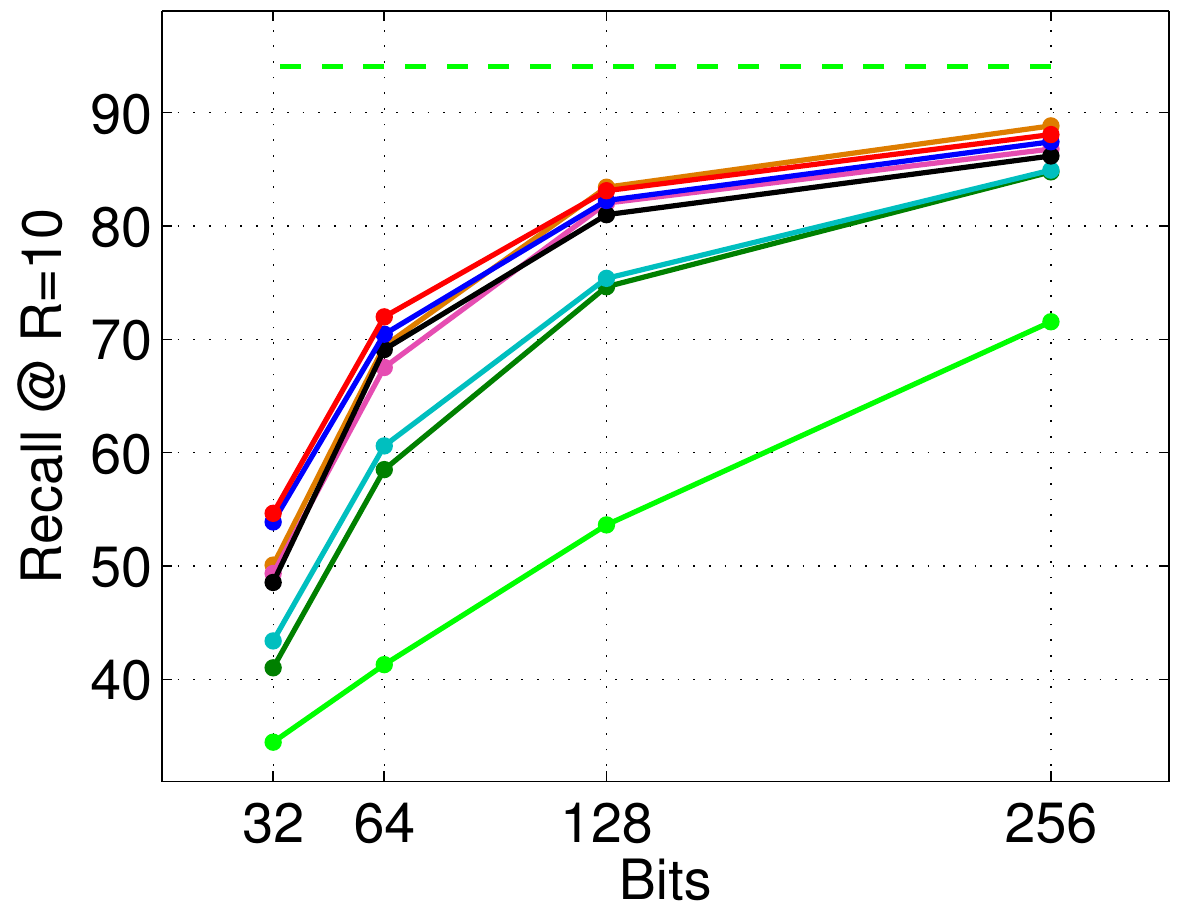} &
			\includegraphics[width=2in]{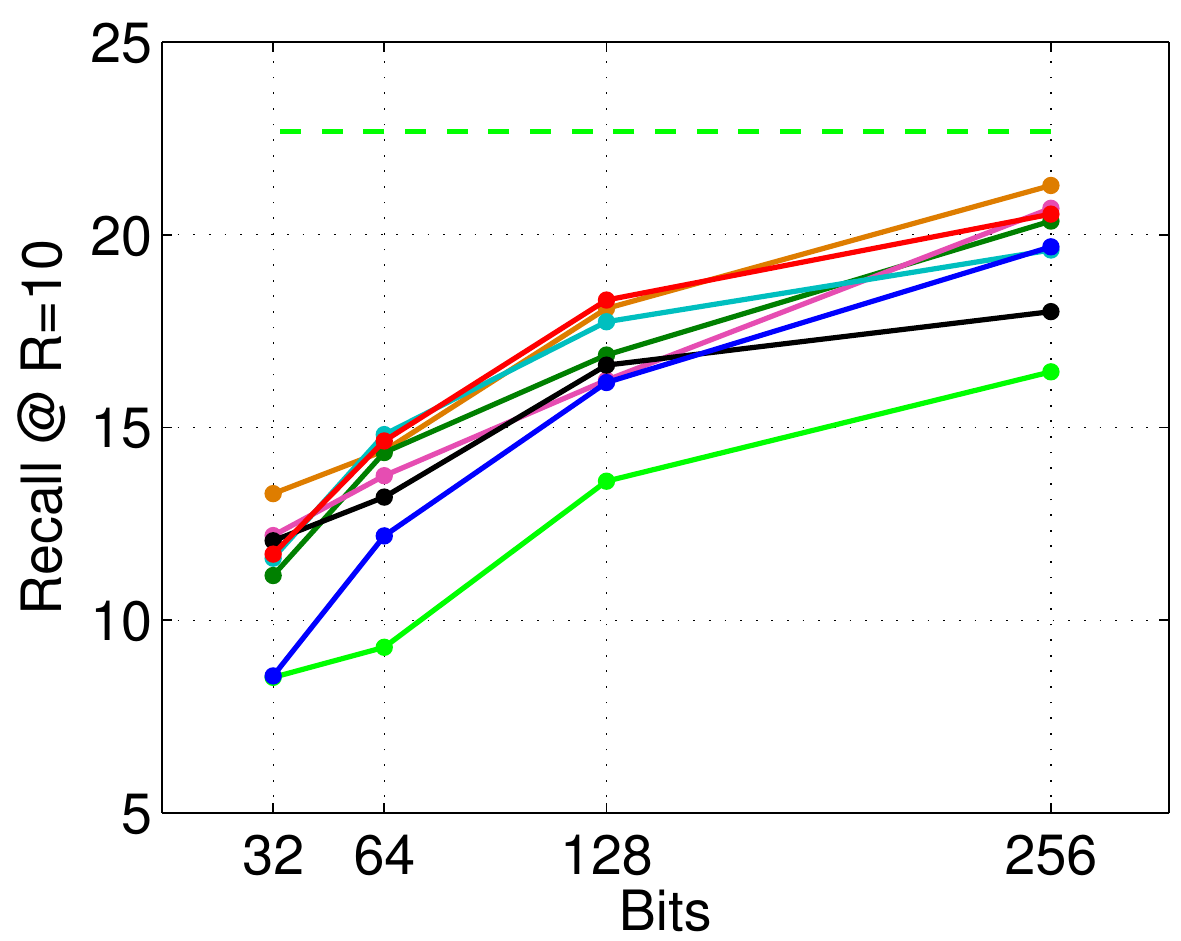} \\
			{\it \footnotesize (a) Holidays: Recall @ $R=10$} & 
			{\it \footnotesize (b) UKbench:  Recall @ $R=10$} &
			{\it \footnotesize (c) Oxford5k: Recall @ $R=10$} \\
			\includegraphics[width=2in]{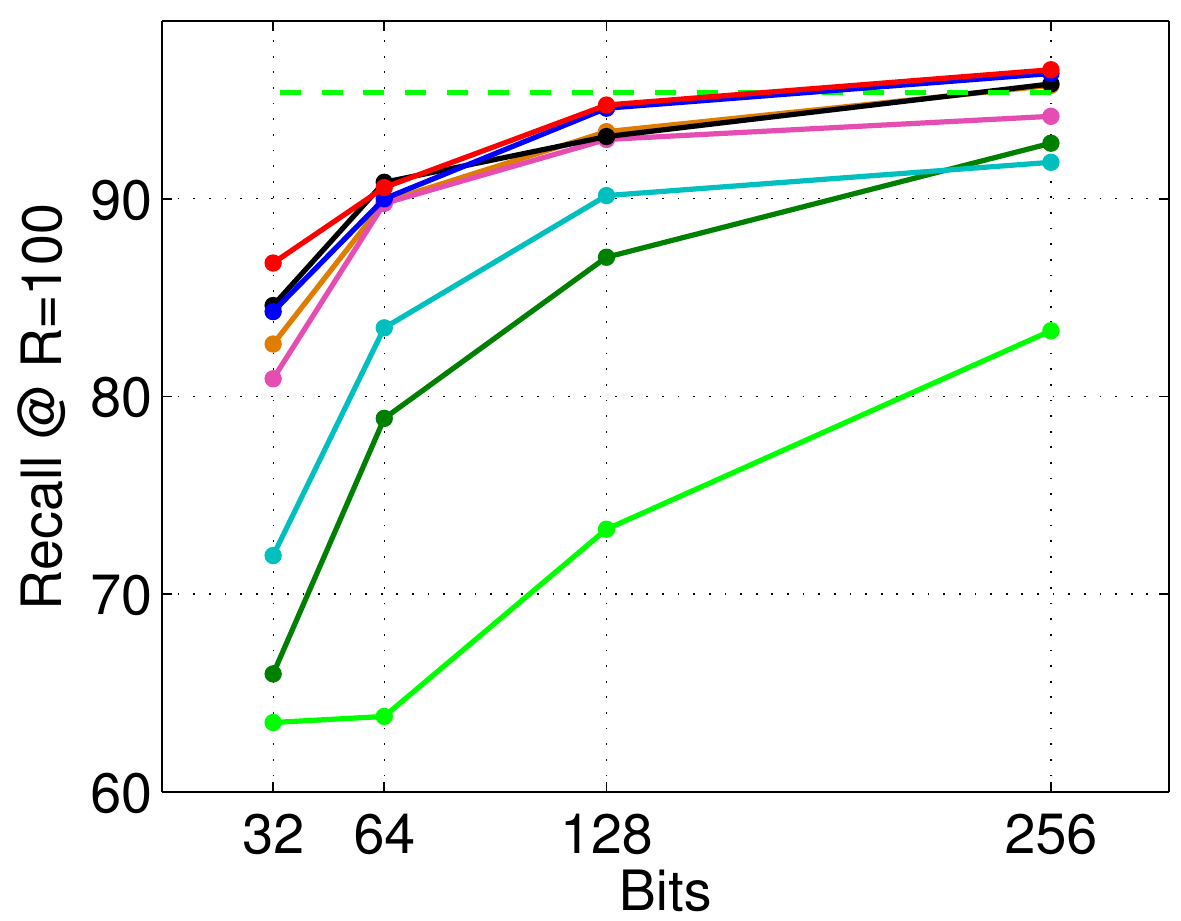} &
			\includegraphics[width=2in]{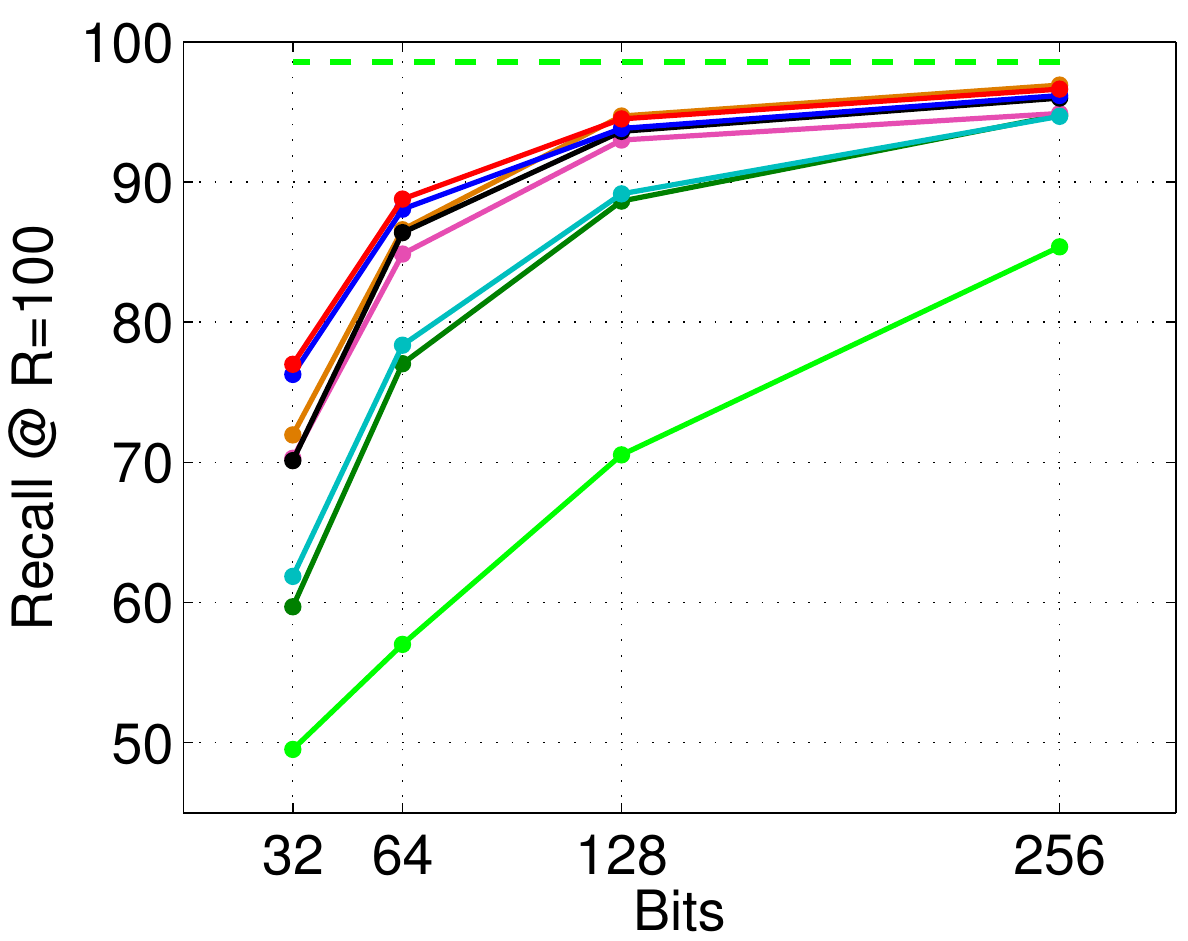} &
			\includegraphics[width=2in]{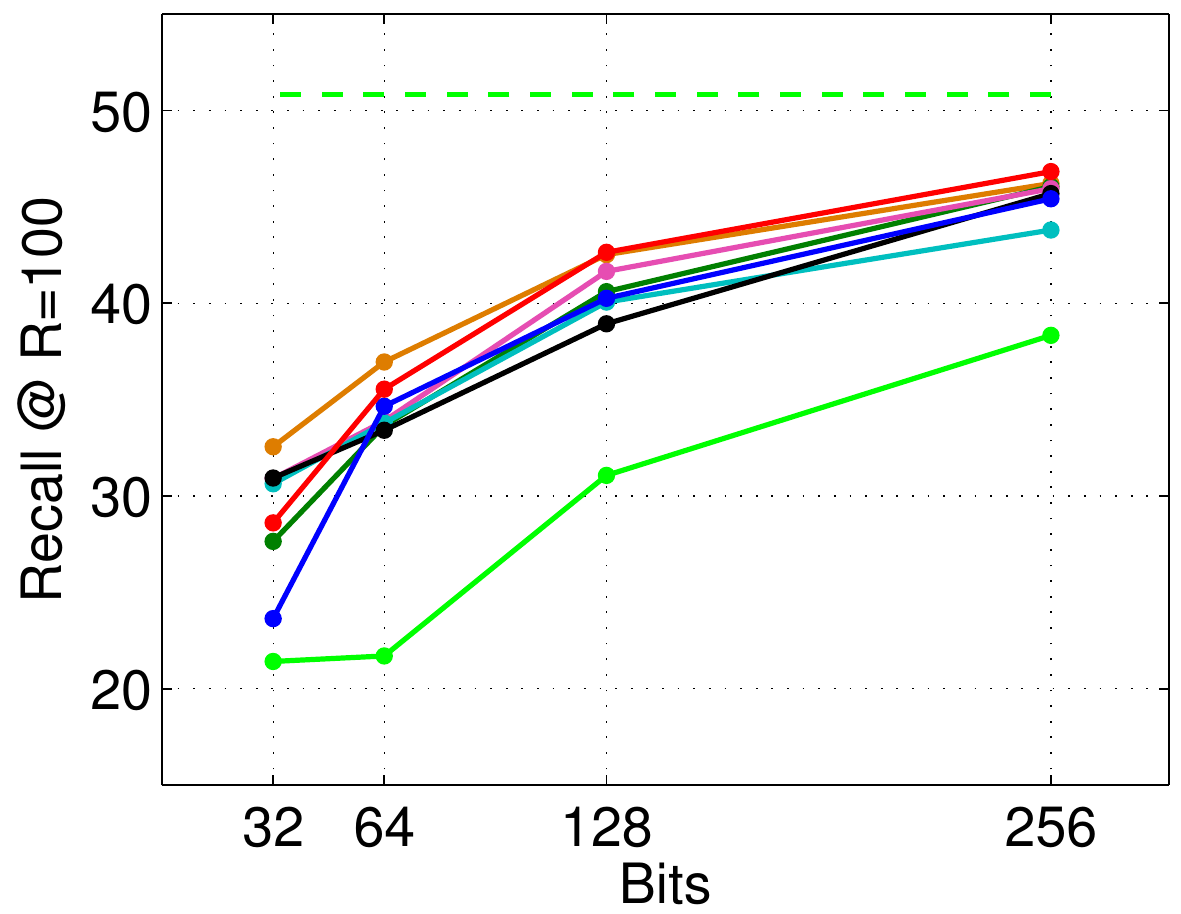} \\
			{\it \footnotesize (d) Holidays: Recall @ $R=100$} & 
			{\it \footnotesize (e) UKbench: Recall @ $R=100$} & 
			{\it \footnotesize (f) Oxford5k: Recall @ $R=100$} \\
			\includegraphics[width=2in]{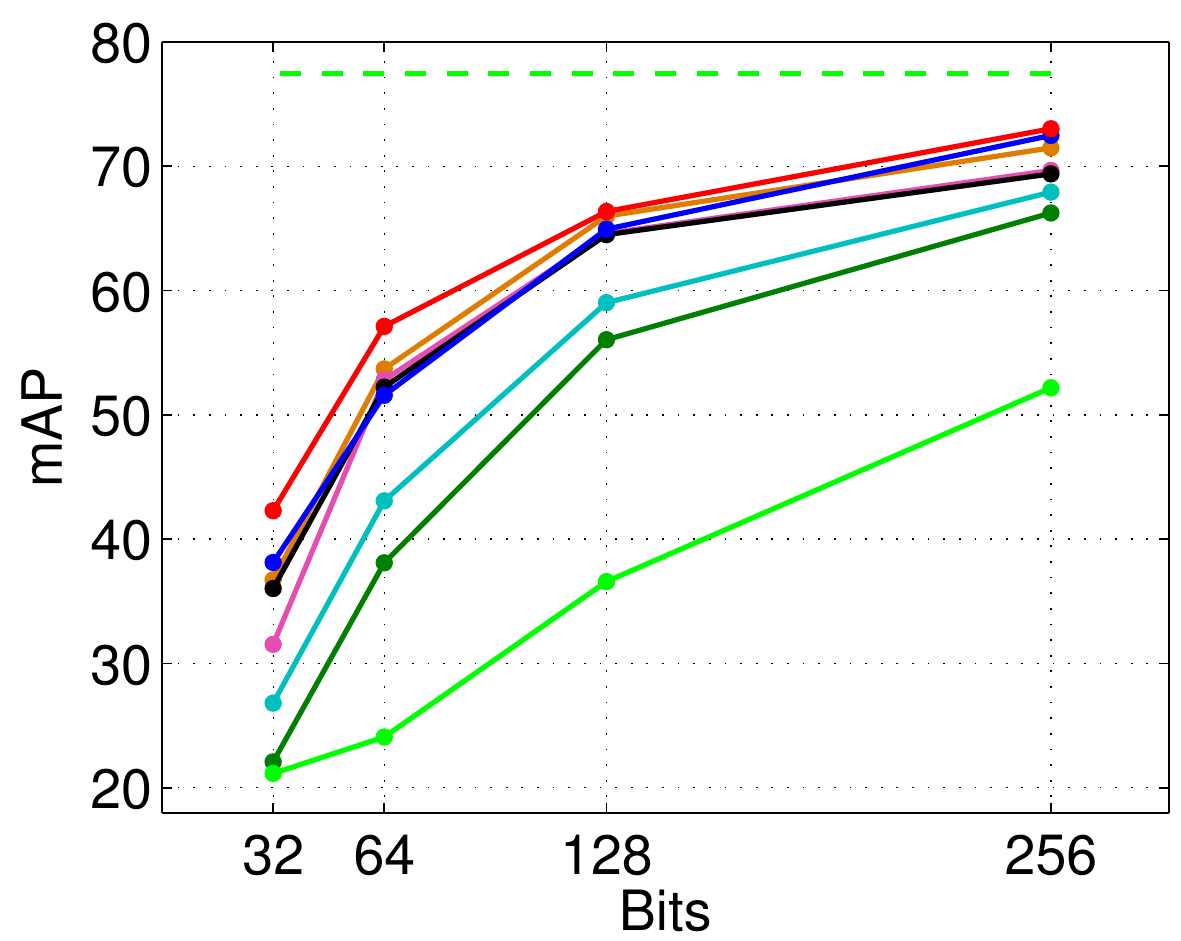} &
			\includegraphics[width=2in]{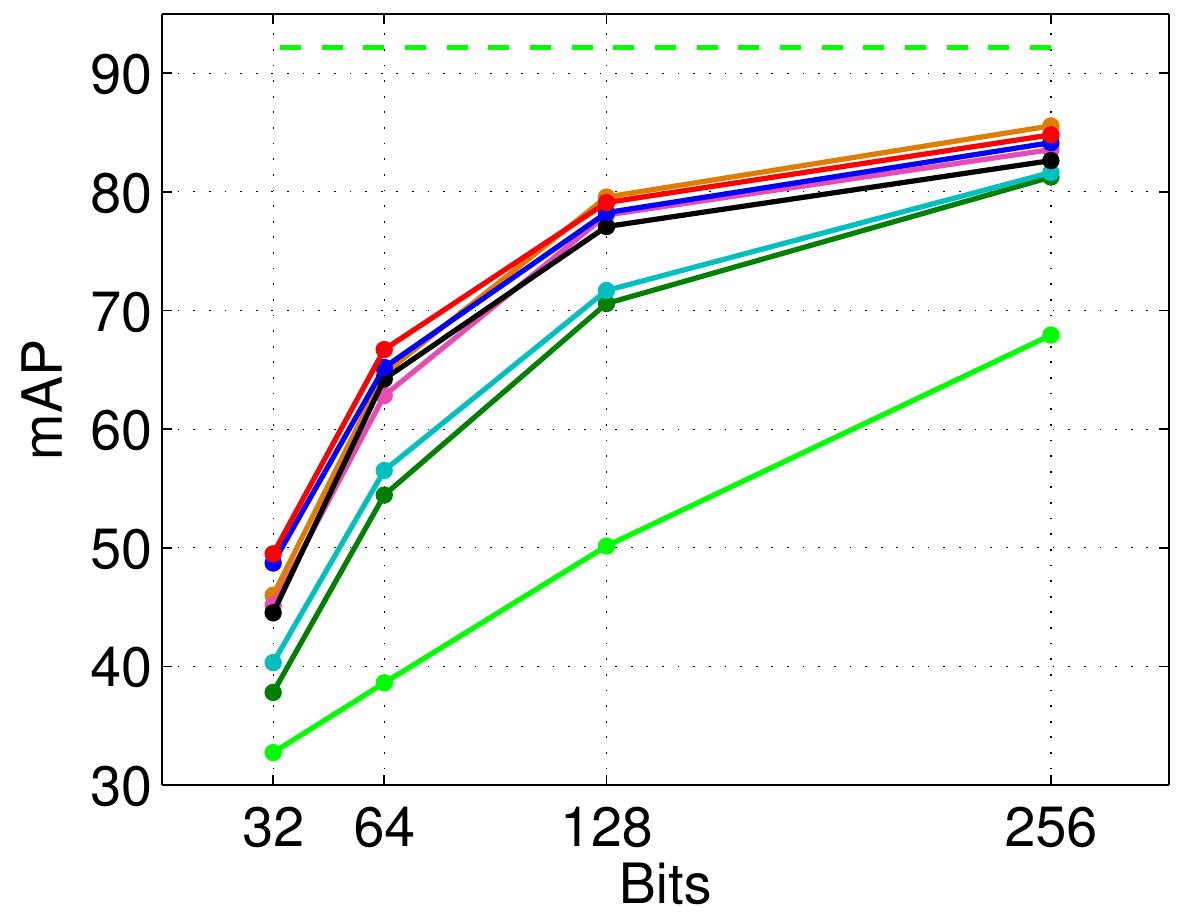} &
			\includegraphics[width=2in]{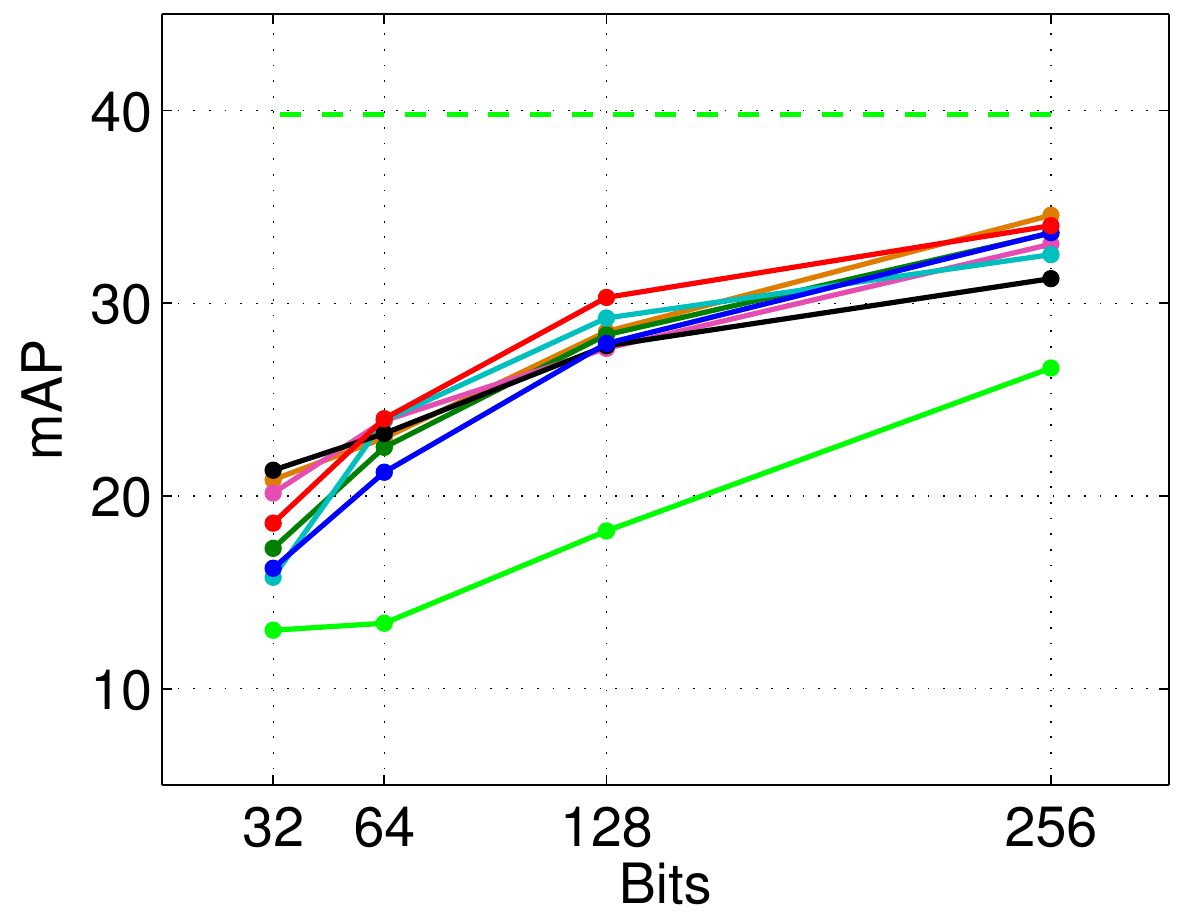} \\
			{\it \footnotesize (g) Holidays: mAP} & 
			{\it \footnotesize (h) UKbench: mAP} & 
			{\it \footnotesize (i) Oxford5k: mAP} \\
		\end{tabular}
		\caption{Retrieval results in terms of Recall @ $R=10$, Recall @ $R=100$ and mAP for different compression schemes on Holidays, UKbench and Oxford5k datasets. 
			The proposed UTH outperforms most schemes across various bitrates on different datasets.}
		\label{fig:retrieval_results}
		}	
\end{figure*}

As a baseline, we also show the performance of the uncompressed descriptors (4096-dimensional floating point representation).
$L_2$ norm is used for uncompressed descriptors, while hamming distances are used for all binary hashing schemes.
We present recall@10, recall@100 and mAP results in Figure~\ref{fig:retrieval_results} for 
the different schemes on Holidays, UKbench and Oxford5k data sets, at varying bitrates.
We make the following observations.

\squishlist

\item With the fine-tuning stage, the proposed {\it UTH} performs much better than SRBM on all datasets, 
especially for Holidays and Oxford5k at low bitrates.
Besides, {\it ITQ}, {\it PCAHash} and {\it SH} obtain higher accuracy than {\it LSH} and {\it BPBC}.
{\it SKLSH} performs the worst.
The ordering of schemes is largely consistent across recall@10, recall@100 and mAP results.

\item {\it UTH} outperforms most schemes across various bitrates on different datasets.
For both Holidays and UKbench, {\it UTH} performs significantly better than the baseline schemes at extremely low bitrates (e.g., $32$),
for example, $+5\%$ in terms of recall$@100$ than {\it ITQ} on UKbench at $32$ bits.
The baseline schemes catch up in performance as bitrate increases, i.e., the improvements of {\it UTH} over the rest schemes become smaller at $256$ bits (expect SKLSH).
We note that {\it UTH} performs slightly worse than {\it ITQ}, {\it PCAHash} and {\it SH} (about $-2\%$ in mAP) on Oxford5k.
Considering there are only $55$ query images from Oxford5k, 
$-2\%$ in mAP means {\it UTH} is only worse than {\it ITQ}, {\it PCAHash} and {\it SH} on one single query on average, which is a relatively small difference.

\item There is a significant gap between the uncompressed descriptor and all the compression schemes at extremely low bitrates on all datasets,
while the performance of compression schemes approach to uncompressed descriptors as bitrate increases to 256 bits.
For instance, there is a $26\%$ drop in recall$@10$ at $32$ bits on Holidays for {\it UTH}, compared to uncompressed descriptors,
the drop is largely reduced to $2\%$ at $256$ bits.

\squishend

\begin{figure*}
	\centering{
	    \includegraphics[width=4in]{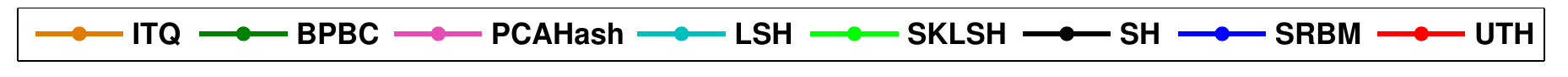}
		\begin{tabular}{@{}c@{} @{}c@{}}
			\includegraphics[width=2.5in]{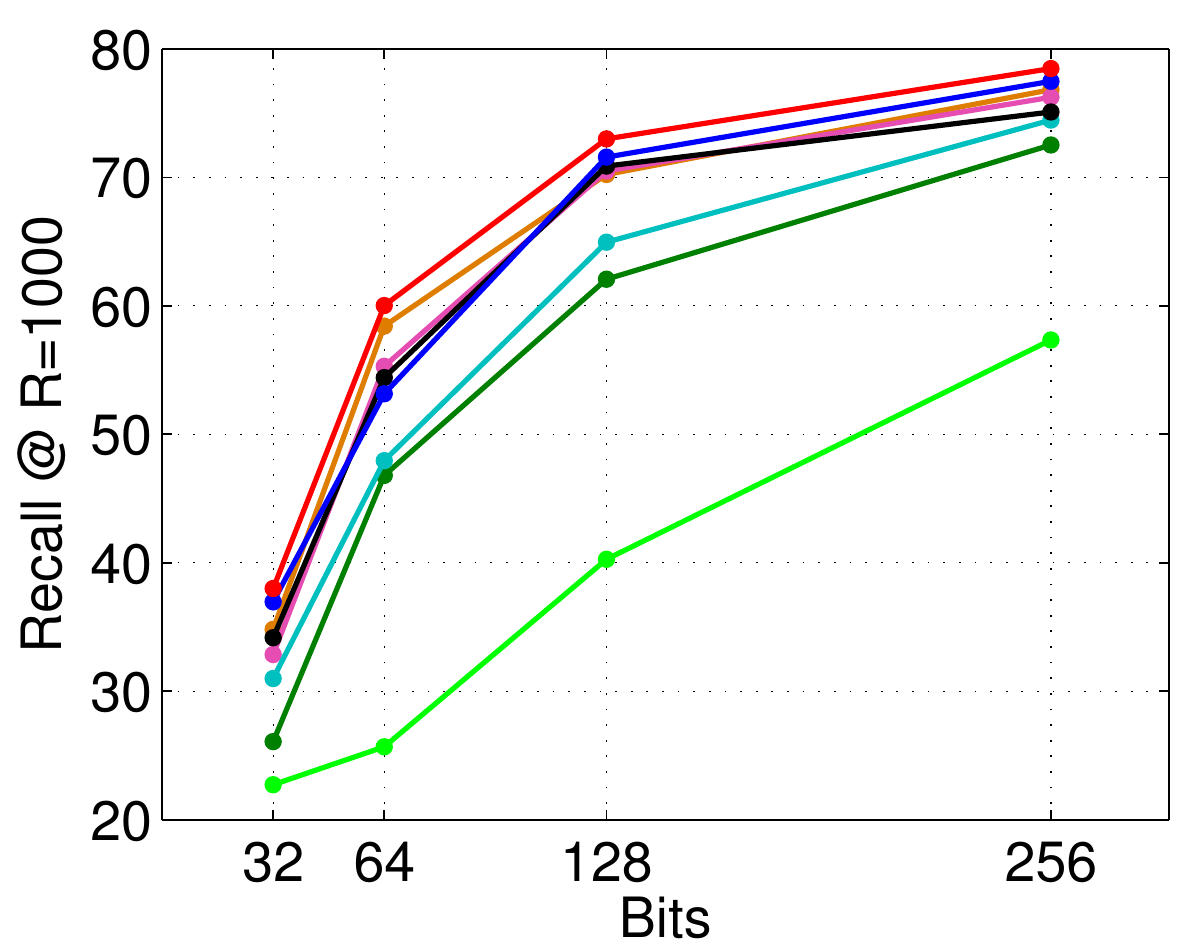} &
			\includegraphics[width=2.5in]{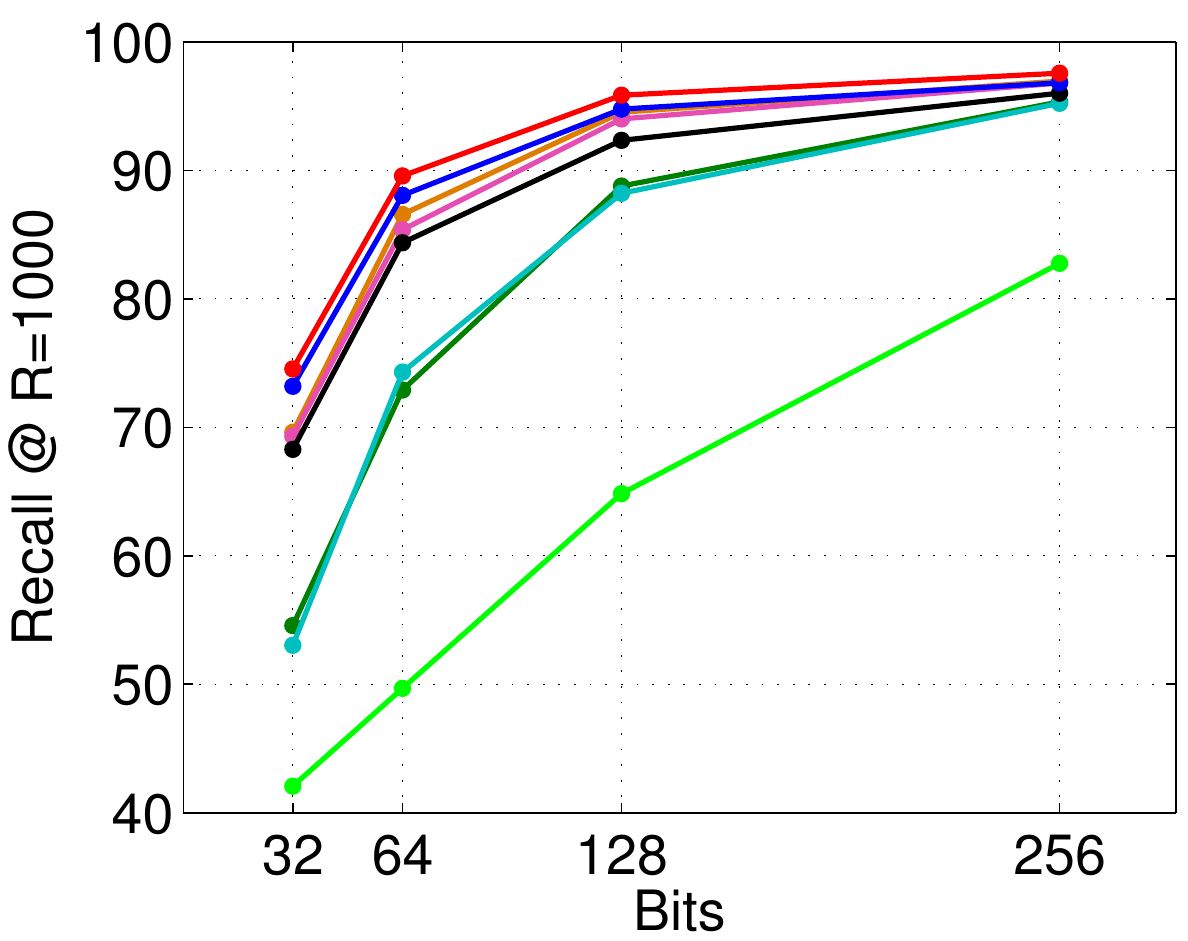} \\		
			{\it \footnotesize (a) Holidays+1M: Recall @ $R=1000$} & 
			{\it \footnotesize (b) UKbench+1M: Recall @ $R=1000$} \\
		\end{tabular}
		\caption{Large scale retrieval results (with 1 million distractor images) for different compression schemes. 
			The proposed UTH outperforms other schemes at all bitrates on both Holidays+1M and UKbench+1M.}
		\label{fig:1m_results}
		}	
\end{figure*}

Finally, we present large-scale retrieval experiments in Figure~\ref{fig:1m_results} for 
{\it Holidays} (500 queries) and {\it UKbench} (10200 queries) data sets combined with 1 million distractor images, respectively.
For instance retrieval, the GCC step is computationally complex and can only be performed on a small number of images.
As a result, it is important for the relevant image to be present in the short list, so that the GCC step can find it.
Hence, we present recall at typical operating points $R=1000$ after the first step in the retrieval pipeline: matching of global descriptors.
Best parameters for {\it UTH} are chosen as described before.
We note that {\it UTH} outperforms all other schemes at all bitrates.
This trend are consistent with the small scale experiments.

Considering the significant drop in performance at 32 bits,
improving retrieval accuracy at extremely low rates is an exciting direction for future work.
Besides, the deep ranking scheme may be further improved using supervised information: 
e.g., taking into side information such as image labels to find more compact and discriminative embedding functions.
Another promising direction would be to learn compact global descriptors for instance retrieval, 
directly from image pixel data using CNNs~\cite{VeryDeepNeuralNets}.

\section{Conclusion}

We proposed UTH, a novel fully-unsupervised method for compressing global image descriptors to extremely small binary hashes (32-256 bits).
Following a pre-training step using SRBMs, 
the model parameters are fine-tuned using image triplets in order to preserve the good retrieval performances of the uncompressed descriptors.
With a thorough empirical evaluation, 
we showed that the fine-tuning step consistently improves the retrieval results and that UTH comes close to matching the performance of the uncompressed descriptor at 256 bits.
On average, results suggest that UTH is currently best unsupervised hashing scheme outperforming other popular schemes such as LSH, PCAHash or ITQ.

\let\oldthebibliography=\thebibliography
  \let\endoldthebibliography=\endthebibliography
  \renewenvironment{thebibliography}[1]{%
    \begin{oldthebibliography}{#1}%
      \setlength{\parskip}{0ex}%
      \setlength{\itemsep}{0ex}%
  }%
  {%
    \end{oldthebibliography}%
  }

{
\scriptsize{
\bibliography{dcc2015_arxiv} 
\bibliographystyle{IEEEtran}
}}

\end{document}